# High-quality single InGaAs/GaAs quantum dot growth on a CMOS-compatible silicon substrate for quantum photonic applications

**Imad Limame,[1] Peter Ludewig,[2] Ching-Wen Shih,[1] Marcel Hohn,[1] Chirag C. Palekar,[1] Wolfgang Stolz,[2] AND Stephan Reitzenstein[1,*]**

[1] *Institute for Solid State Physics, Technical University of Berlin, Hardenbergstraße 36, D-10623 Berlin, Germany*
[2] *Structure and Technology Research Laboratory, Philipps University of Marburg, Hans-Meerwein-Straße 6, D-35032 Marburg, Germany*
*stephan.reitzenstein@physik.tu-berlin.de*

**Abstract:** We present the direct heteroepitaxial growth of high-quality InGaAs quantum dots on silicon, enabling scalable, cost-effective quantum photonics devices compatible with CMOS technology. GaAs heterostructures are grown on silicon via a GaP buffer and defect-reducing layers. These epitaxial quantum dots exhibit optical properties akin to those on traditional GaAs substrates, promising vast potential for the heteroepitaxy approach. They demonstrate strong multi-photon suppression with $g^{(2)}(\tau) = (3.7 \pm 0.2) \cdot 10^{-2}$ and high photon indistinguishability $V = (66 \pm 19)\%$ under non-resonance excitation. We achieve up to $(18 \pm 1)\%$ photon extraction efficiency with a backside distributed Bragg mirror, marking a crucial step toward silicon-based quantum nanophotonics.

## 1. Introduction

The integration of classical optically active materials onto the silicon (Si) platform has garnered significant interest in both scientific and industrial communities for the last two decades. While Si is widely known for being cost-effective and is extensively used in semiconductor technology, its indirect bandgap hinders its potential for optoelectronic applications. One route for the fabrication of Si-compatible optoelectronic devices is based on rather complex wafer-bonding techniques. Alternatively, the direct growth of III-V materials, which have a direct bandgap and exhibit excellent optical properties, on Si is very attractive. However, this approach has proven to be a challenging endeavor due to various factors such as lattice mismatch, differences in thermal expansion coefficient, chemical reactivity at the Si surface, and the formation of dislocations. Nevertheless, considerable progress has been made to address some of those challenges. For instance, buffer layers and graded compositions can be used to gradually accommodate the lattice mismatch and reduce the formation of dislocations [1,2]. These advancements have led to groundbreaking achievements in Si-based optoelectronics, such as room-temperature emitting diodes and lasers directly grown on Si, among other notable developments [3–5]: Since the first laser in 2003 to the most recent optical devices on Si, the realization of integrated photonic circuits in conjunction with the well-established complementary metal oxide semiconductor (CMOS) fabrication technology has been a highly sought-after goal [6]. In particular, such a technology is interesting for low-cost on-chip interconnect systems, high-speed data communication, and computing technologies [7–9].

Despite significant advances in conventional silicon photonics using classical light sources, progress in silicon-compatible quantum photonics has been slow and hindered by the challenge of achieving direct and high-quality growth of single quantum emitters on the Si platform [10,11]. Although there have been advancements in the post-growth integration of QDs on Si, the complexity of the process limits the scalability and cost-effectiveness [12]. Silicon quantum photonics aims to leverage the unique properties of quantum systems, such as superposition, entanglement, and photon indistinguishability, to enable a cost-effective combination of state-of-the-art Si electronics and advanced quantum photonics. A particularly attractive avenue of research in the field of single-photon sources (SPS) focuses on the Stranski–Krastanov (SK) growth of InGaAs/GaAs quantum dots (QDs) on GaAs or InP substrates. Such QDs have demonstrated close-to-ideal quantum properties including high single-photon purity, high entanglement fidelity, and low distinguishability combined with high extraction efficiency, while embedded in photonic structures and spanning the wavelength range from 900 nm up to the telecom C-band [13–16]. The exceptional quantum optical properties of these emitters have garnered considerable attention in quantum information technology [17], including quantum key distribution (QKD) [18,19], and boson sampling [20]. In addition to the excellent quantum optical properties of QDs, several techniques have been established to deterministically realize QD-based quantum photonic devices, paving the way for the scalable fabrication of QD-based optoelectronic systems. Deterministic nanofabrication platforms like the site-controlled growth based on a buried stressor [21], and nanohole arrays [22] or post-growth selective integration using in-situ lithography and marker-based lithography have been developed, allowing for precise control over the placement of the QDs and facilitating the creation of complex, integrated quantum devices. [21,23,24]

In this work, we present the growth of high-quality, low-density InGaAs/GaAs QDs on an exactly oriented ($\pm 0{,}5°$) Si (001) substrate. We achieved this by utilizing a GaP buffer layer as the initial step for the subsequent growth of III/V materials on Si. This innovative technique has been investigated in previous reports [25, 26], and is further optimized in the present work to grow QDs with low surface density and unpreceded emission properties on Si substrate. The grown quantum emitters exhibit excellent optical characteristics comparable to InGaAs/GaAs QDs based on GaAs substrates. Under pulsed non-resonant wetting-layer excitation at 930 nm, we obtain high multi-photon suppression, $g^{(2)}(0) = (3.7 \pm 0.2) \cdot 10^{-2}$ (post-selected: $g^{(2)}_{Post} = (3.5 \pm 0.2) \cdot 10^{-3}$) and achieve the emission of indistinguishable photons with a visibility of $(65.6 \pm 19.1)\%$. Additionally, the photon extraction efficiency, reaching up to $(18 \pm 1)$ % for a simple planar structure with a backside distribute Bragg reflector, aligns closely with the predictions from finite element method (FEM) simulations. By combining the excellent quantum optical properties of InGaAs/GaAs QDs, the available techniques for deterministic fabrication of quantum photonic devices, the attractive waveguiding properties of Si, and CMOS fabrication technology, our results can drive forward the development of cost-effective and scalable Si quantum optoelectronics.

## 2. Fabrication and methods

The semiconductor heterostructure is grown in two epitaxial steps using metal-organic chemical vapor deposition with the schematic of the investigated structure being depicted in Fig. 1. (a). The GaP and the AlGaAs/Gas relaxation buffer were deposited on an exactly oriented 300 mm Si (001) wafer in an AIXTRON Cruis R CCS twin reactor system at NAsP III/V GmbH. We used our established GaP-on-Si template process with a 5 µm thick Si:P buffer. More detailed information about the anti-phase domain-free deposition of GaP on exact Si (001) substrates can be found in [25, 26]. The 50 nm GaAs-LT nucleation layer was deposited at 450 °C onto the GaP-on-Si template. The wafer was then heated to 600 °C under TBAs (tertiarybutylarsine) stabilization and additional 50 nm GaAs:Si were grown. Subsequently, 100 nm AlGaAs:Si, 50 nm GaAs:Si, 100 nm GaInP:Si and 2.5 µm GaAs:Si were deposited for defect filtering and relaxation of GaAs lattice constant. The 300 mm wafer was then cleaved into smaller-sized pieces of 4 cm x 4 cm and transferred to the AIX 200/4 of the Technical University of Berlin for QD growth. During the second epitaxial step, the template substrate is heated to 735 °C to remove any native oxide. To ensure high crystal quality, a 300 nm undoped, and high V/III-ratio (200) GaAs layer is deposited. Subsequently, a distributed Bragg reflector (DBR) is grown, consisting of 33.5 pairs of alternating λ/4-layers of GaAs and $Al_{0.90}Ga_{0.10}As$ with nominal thicknesses of 68.1 nm and 80.6 nm, respectively. This DBR is included as backside mirror to enhance the outcoupling efficiency of the QDs in the active layer. Following the DBR section, 137 nm of GaAs is deposited as half of a λ/4 cavity. The sample is then cooled down to 500 °C, and a 0.40 thin $In_{0.33}Ga_{0.67}As$ wetting layer (WL) is deposited. Following a growth interruption of about 40 seconds, the QDs with a density of approximately $1 \cdot 10^9$ $cm^{-2}$ are capped by 0.8 nm of GaAs, and the sample is heated up to 615 °C to desorb the remaining indium from the surface. Finally, a 135 nm thick GaAs layer is grown to protect the emitter from any surface states. A scanning electron microscopy (SEM) image of the cleaved final structure is presented in Fig. 1. (b), showing a side view of each grown layer, from the Si wafer at the bottom to the final GaAs capping layer on the top, where the QDs are located. Surface morphology analysis is conducted using atomic force microscopy (AFM). Fig. 1. (c) and Fig. S1. (a) depict a 90 µm x 90 µm and a zoomed-in 10 x 10 µm AFM scan, respectively. The structure under investigation displays a rough macroscopic surface, featuring grains ranging from 6 to 12 µm in lateral size and up to 400 nm in height. Those grain-like features stem from the three-dimensional nucleation of GaP on the Si substrate [25]. Additionally, the coalescing of GaP results in stacking faults, a phenomenon evident in Fig. S1. (a). We also observe the presence of misfit dislocations, as shown in Fig. S1. (b), which form at the Si/GaP interface to accommodate for the lattice mismatch. These defects contribute to the degradation of optical and electrical characteristics, ultimately limiting the performance of conventional optoelectronic devices on the macroscopic level [25, 26].

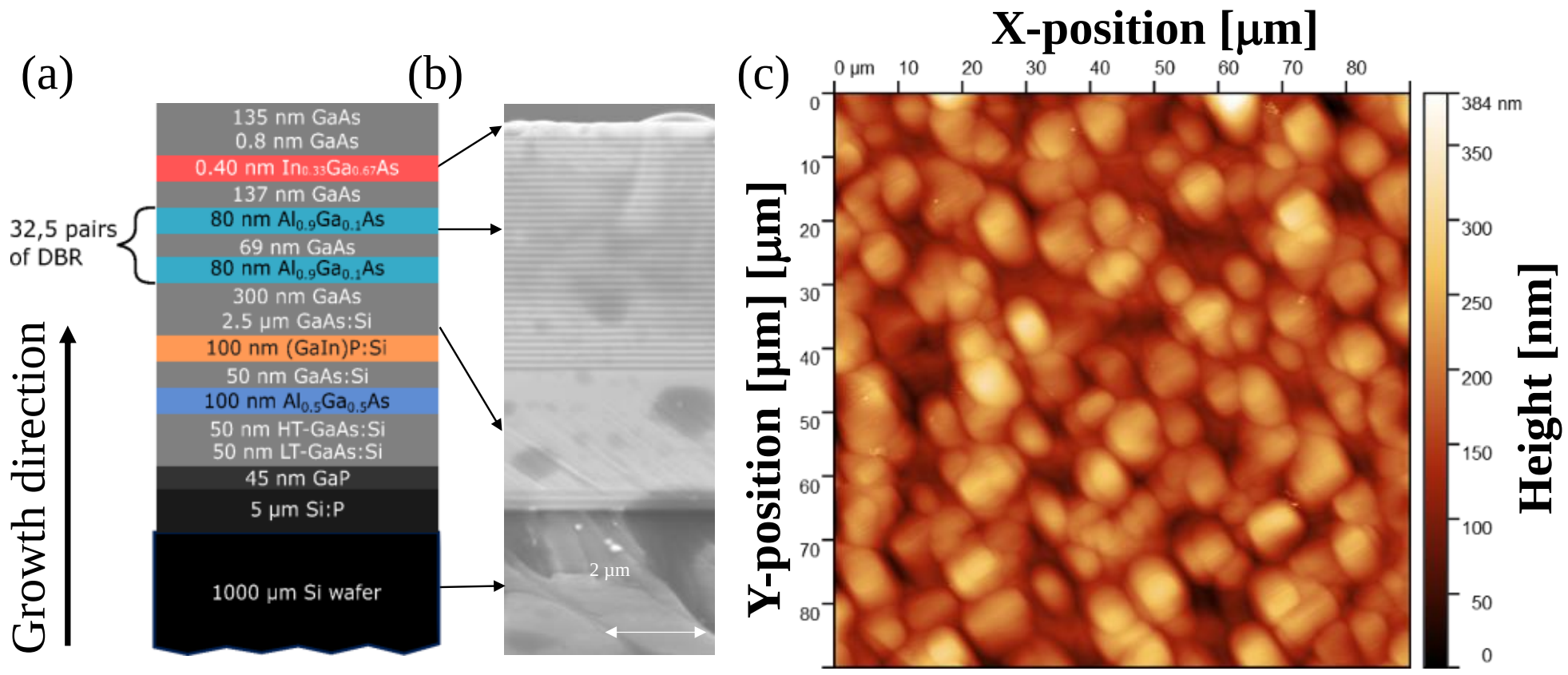

Fig. 1. (a) Layer design of the investigated structure grown on a 1 mm thick undoped Si wafer. 32.5 pairs of a GaAs/$Al_{0.90}Ga_{0.10}As$ DBR to enhance the photon extraction efficiency of the QDs is deposited at 700 °C and a 0.40 nm thin $In_{0.33}Ga_{0.67}As$ wetting layer is grown at a lower temperature of 500 °C and capped by 135 nm of GaAs. (b) A crosssection SEM image of the final structure showing the Si wafer, DBR, and top a GaAs capping layer. (c) 90 µm x 90 µm AFM image of the top surface of the final structure.

The quantum optical characterization of the silicon-compatible III/V-heterostructures is conducted at 4.2 K using a micro-photoluminescence (µPL) setup comprising a closed-cycle cryostat, as illustrated in Fig. S2. To excite the QD of interest, a 2 ps pulsed tuneable optical parametric oscillator (OPO) laser at a repetition frequency of $(80.77 \pm 0.4)$ MHz is focused on the sample using a 0.8 NA aspheric lens, which also collects the emitted photons. We apply a non-resonant excitation scheme into the wetting layer of the QDs at 930 nm. A 950 nm long pass (LP), slightly tilted to adjust the cutoff wavelength, filters the pump laser and prevents stray light in the QD collection of the QD emission. The collected light is then guided in a spectrometer equipped with a allowing a resolution of 30 µeV via a 1200 lines/mm grating. Alternatively, the signal can be collected behind the monochromator into a single-mode fiber. In the Hanbury Brown and Twiss (HBT) setup, the signal is split equally by a fiber-based 50:50 splitter and directed to two superconducting nanowire single-photon detectors (SNSPDs) with a combined time resolution of 80 ps. The output of the SNSPD is then correlated using a time-correlated single-photon counter (TCSPC) to measure the second-order autocorrelation function. For the lifetime measurement, a single channel is selected, and the $(80.77 \pm 0.4)$ MHz pulsed laser serves as the start/stop trigger signal. The indistinguishability of the emitted photons is measured in a Hong-Ou-Mandel (HOM) interferometer. To achieve this, a 4 ns delay line is introduced in the excitation path. Similar to the HBT setup, the signal collected after the monochromator is guided into a single-mode fiber. After exiting the fiber, photons are linearly polarized, and a combination of a λ/2 plate and a polarizing beamsplitter (PBS) is used to achieve a precise 50/50 splitting ratio. Both arms of the interferometer are equipped with λ/2 and λ/4 plates to precisely control the polarization. In one of the branches, a second 4 ns delay line is inserted to achieve a simultaneous impinge of two heralding photons. The HOM interference is then measured at the final 50/50 beamsplitter, where the correlation can be detected in co- and cross-polarization.

## 3. Results and Discussion

To investigate the quantum optical properties of the Si-grown InGaAs/GaAs QDs, a bright emitter with narrow emission linewidth is selected. Fig. 2. (a) shows a waterfall plot of the µPL emission spectrum of the investigated QD at excitation powers ranging from 10 to 270 µW. At a pump power of 110 µW, the single QD exhibits a pronounced excitonic emission line at 1.3015 eV (approximately 950 nm) with a resolution-limited linewidth of 30 µeV. To identify the excitonic complexes of the QD system, excitation power-dependent, and polarization-resolved µPL measurements are conducted (shown in Fig. S3). Based on these measurements, we identify three transitions associated with the investigated QD: two neutral excitonic states and one charged state, as discerned from the degree of linear polarization (DLP) presented in Fig. 2. (b), blue curve. The neutral exciton $(X)$ at 1.3067 eV has a linear power dependence $(I_X \sim P^{1.15})$ and exhibits a fine structure splitting (FSS) of $(30.5 \pm 0.1)$ µeV, see 0° and 90° polarized spectra in black and red, respectively in the inset of Fig. 2(b). The extracted FSS value is comparable with values reported previously for similar InGaAs/GaAs QDs grown on GaAs, which is typical for this material system. The corresponding biexciton $(XX)$ at 1.3042 eV is recognizable by its opposite linear polarization if compared to the exciton in Fig. S3. (b), along with the nearly quadratic power dependency $(I_X \sim P^{2.14})$, as depicted in Fig. S3. (a). Lastly, the charged biexciton $(XX^{+/-})$ with an energy of 1.3015 eV and a (resultion limited) linewidth of 31 µeV is identified by the absence of FSS in the polarization-resolved measurement. It exhibits a super-linear intensity dependence on the excitation power $(I_X \sim P^{2.48})$, as presented in Fig. S3. Moreover, the emergence of the aforementioned state (XX+/-) at higher pump powers relative to the exciton (X) and similar excitations as the biexciton (XX) provides additional validation for the nature of the transition.

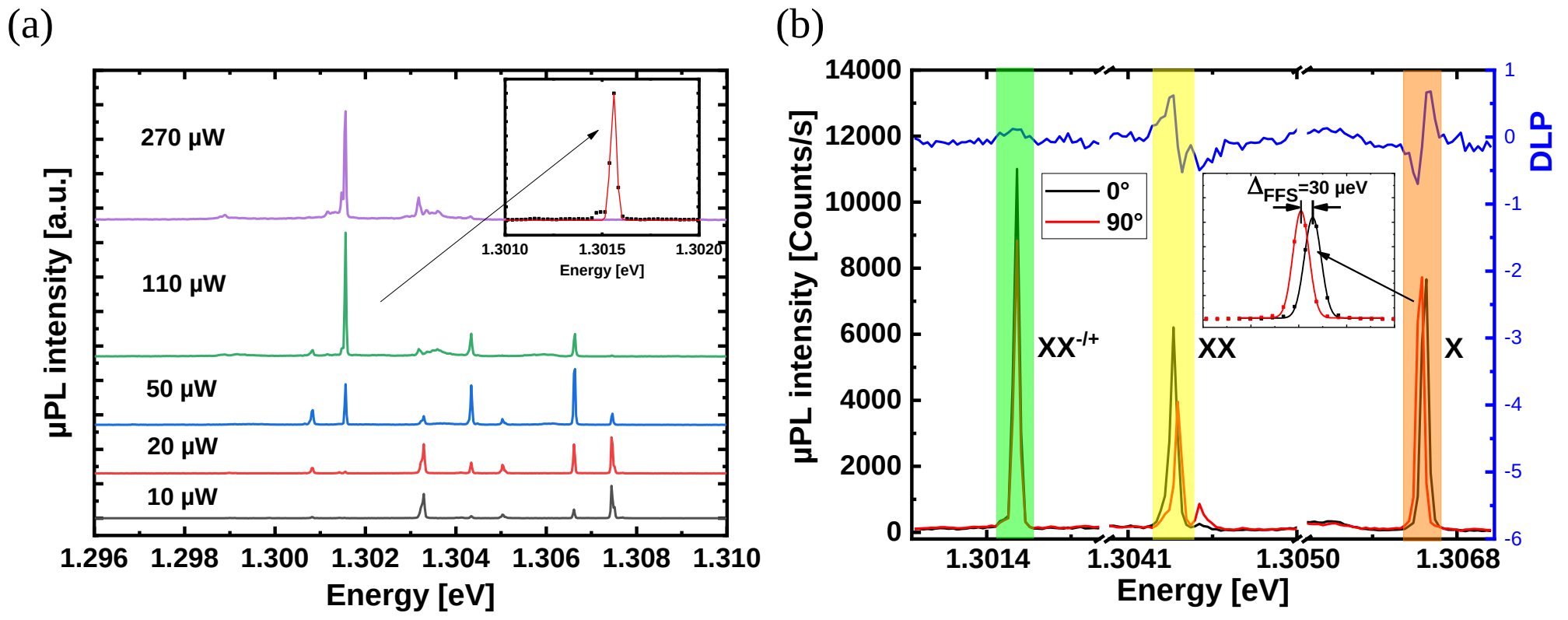

Fig. 2. (a) Waterfall plot of the µPL emission of the investigated InGaAs/GaAs QD grown on Si under 930 nm pulsed excitation with pump powers ranging from 10 to 270 µW at 4.2 K. The inset shows a Gaussian fit of the charged biexciton line at 1.3014 eV yielding a resolution limited linewidth of 30 µeV. (b) µPL spectra of the exciton (1.3064 eV), biexciton (1.3042 eV), and charged biexciton (1.3015 eV) at a pump power of 110 µW for 0° and 90° polarization angle highlighted in orange, yellow, and green, respectively. Zoom-in view of the X line showcasing the fine structure splitting (FSS) of $(30.5 \pm 0.1)$ µeV. The blue curve shows the degree of linear polarization (DLP).

In addition to quantum optical properties, which are discussed in the following sections, the photon extraction efficiency is an important parameter for the practical usage of quantum light sources, because it later influences for instance the communication rate in quantum networks. First, we determine the expected photon extraction efficiency of the given sample design via numerical simulations based on the FEM using the solver JCMsuite by JCMwave. Fig. S4. (a) presents the resulting photon extraction efficiency as a function of emission wavelength for 0.8 NA. At the wavelength of the considered QD (941.43 nm), the simulation yields an extraction efficiency of 18.67 %. Assuming a quantum efficiency of the QD equal to 1 and a QD occupation probability of 1 (achieved usually in saturation of the neutral exciton) the experimental photon extraction efficiency is determined as follows:

$$\eta_{\mathrm{QD}} = \frac{\Gamma_{\mathrm{QD}}}{\Gamma_{\mathrm{laser}} \cdot \eta_{\mathrm{setup}}} \quad . \qquad (1)$$

Here, $\Gamma_{\mathrm{QD}}$ denotes the count rate of QD emission at the single-photon detector, divided by the pulsed laser repetition rate of $\Gamma_{\mathrm{laser}} = (80.77 \pm 0.4)$ MHz and the setup efficiency $\eta_{\mathrm{setup}}$, which is $(4.29 \pm 0.03)$ % in the present case. Under pulsed excitation at 800 nm and 20 µW, we measured $\Gamma_{\mathrm{QD}} = (636 \pm 34)$ kcounts/s at saturation as can be seen in Fig. S4. (b), leading to an extraction efficiency of $(18 \pm 1)$ % for this QD. Notably, the good agreement between the theoretical prediction and the experimentally obtained photon extraction efficiency indicates a high quantum efficiency of the QD which underlines the optical high quality of the QD (and DBRs) grown on Si substrate.

Fig. 3(a) depicts a time-resolved measurement of the QD's charged biexcitonic state after spectrally filtering the $XX^{(-/+)}$ line through the monochromator. A linear fitting in the logarithmic plot reveals a lifetime $\tau$ of $(1.55 \pm 0.01)$ ns, as shown in Fig. 3(a). This lifetime is compatible with typical vales reported for SK-grown InGaAs/GaAs QDs on GaAs substrates. Next, to study the quantum optical quality of the grown QDs on Si substrate, we determine the multi-photon emission by measuring the pulsed second-order photon correlation function using a standard HBT setup under pulsed non-resonant wetting-layer (930 nm) excitation. As shown in Fig. 2. (b), pronounced antibunching at $t = 0$ is observed at a pump power of 38 µW, indicating strong multi-photon suppression. The corresponding $g^{(2)}_{Area}(0)$ value, calculated by dividing the area of the peak at zero time delay by the average area of all the other peaks, is $(2.28 \pm 0.03) \cdot 10^{-2}$. In addition to this integrated $g^{(2)}$-value, we calculated a post-selected single multi-photon suppressoin at $\tau = 0$ (as well as a post-selected HOM visibility below) by fitting the following function to the correlation histograms [27,28]:

$$g^{(2)}(\tau) = A\left[e^{\left(-\frac{|\tau|}{\tau_{dec}}\right)} - e^{\left(-\frac{|\tau|}{\tau_{cap}}\right)}\right] + B\sum_{n\neq 0} e^{\left(-\frac{|\tau - n\tau_0|}{\tau_{dec}}\right)} \qquad (2)$$

With $A$ being a scaling parameter related to the secondary photons emission, $\tau$ the delay time, $\tau_{dec}$ the decay time, $n$ and $B$ being the number and average height of the peaks, respectively. We attribute the pronounced dip observed in both $g^{(2)}(\tau)$ correlation function and the co-polarized HOM measurement at $\tau = 0$ (see inset Fig. 3(b) and Fig. 4(b)) to a well-documented recapture effect of charge carrier, followed by subsequent photon emission [28,29]. This phenomenon is a consequence of non-resonant excitation, which leads to the generation of charge carriers that either remain within the WL or become captured by charge traps. Those states subsequently repopulate the QD after the initial photon emission, exhibiting a characteristic capture time $\tau_{cap}$ and emitting a photon through exciton recombination with a decay time $\tau_{dec}$. The fitting of the second-order photon correlation function for the investigated QD yielded a $\tau_{cap} = (0.18 \pm 0.02)$ ns, $\tau_{dec} = (0.83 \pm 0.01)$ ns, and a post-selected $g^{(2)}_{Post} = (3.5 \pm 0.2) \cdot 10^{-3}$. Based on this analysis, we expect both the $g^{(2)}(0)$ and visibility to improve while resonantly exciting the studied QD to reduce the impact of the recapture process. Nevertheless, the reported $g^{(2)}_{Area}(0)$ value here is better than the only known $g^{(2)}(0)$-value single QDs on Si by approximately one order of magnitude and is comparable to high-quality QDs grown on a smooth defect-free GaAs surface [11]. Furthermore, the measured multi-photon suppression is comparable to state-of-the-art strain-free droplet-etched GaAs QDs [30]. Our findings also indicate that the defects propagating from the growth of GaP on Si do not degrade the optical properties of single QDs. This showcases the enormous potential of single InGaAs/GaAs QDs grown on the GaP/Si substrate for applications in quantum photonics combined with Si electronics.

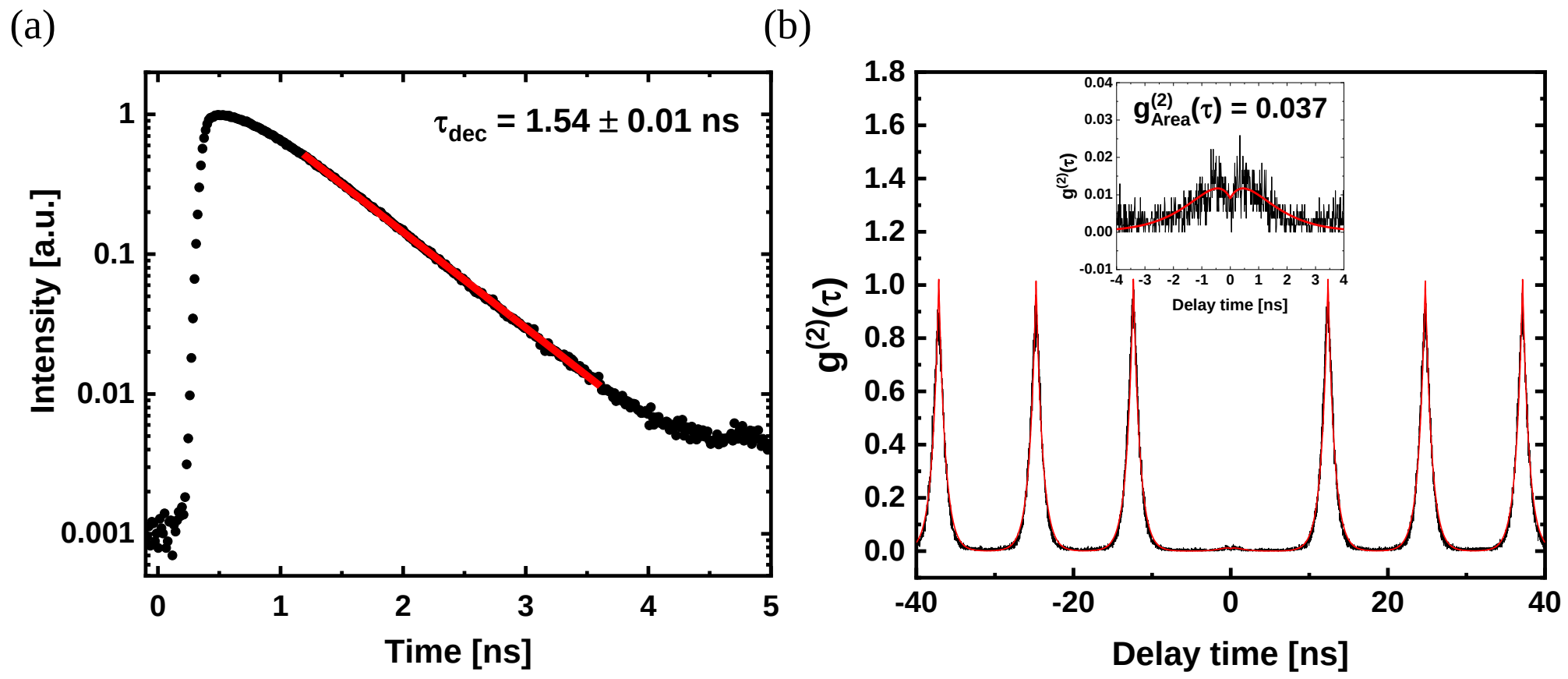

Fig. 3. (a) Time-resolved measurement (black data points) and linear fit (red solid line) of the $E_{XX^{+/-}} = 1.3014$ eV line at a pump power of 34 µW. The QD is excited with an $(80.77 \pm 0.4)$ MHz pulsed laser at 930 nm. The fit yielded a decay time of $\tau = (1.54 \pm 0.01)\ ns$. (b) The second-order photon correlation function for the investigated QD line at an excitation with a pump power of 38 µW, demonstrates clean single-photon emission with a $g^{(2)}_{Area}(0) = (3.7 \pm 0.2) \cdot 10^{-2}$ and a post-selected value of $(3.5 \pm 0.2) \cdot 10^{-3}$.

High photon indistinguishability constitutes a key requirement of various applications in advanced photonic quantum technologies, such as quantum repeater networks based on entanglement distribution via Bell-state measurements [17]. To determine the photon indistinguishability of the QD under study we performed a HOM experiment under non-resonant pulsed excitation at 930 nm with 80 µW pump power using the setup illustrated in Fig. S2. In the initial measurement, the two signals are cross-polarized to ensure the distinguishability of the two photons (Fig. 4. (a)), followed by a co-polarized measurement (Fig. 4. (b)). The resulting two-photon interference data obtained in cross-polarized and co-polarized configurations are depicted in Fig. 4. As a result of the aforementioned (Eq. 2) fit function, the cross-polarized measurement (black curve) displays a $g^{(2)}_{\perp}(0)$ value close to 0.5 as expected, while the co-polarized measurement (red curve) exhibits pronounced anti-bunching with $g^{(2)}_{\parallel}(0) = 0.024$. Considering the results of the co-polarized and the cross-polarized measurement, we obtain a non-post-selected value of $(65.6 \pm 19.1)$ %, while the post-selected visibility is $V^{HOM}_{Post} = \left(g^{(2)}_{\perp}(0) - g^{(2)}_{\parallel}(0)\right) / g^{(2)}_{\perp}(0) = (93.1 \pm 0.4)$ %, confirming the quantum nature of the emitted photons by the InGaAs/GaAs QDs grown on the GaP/Si platform.

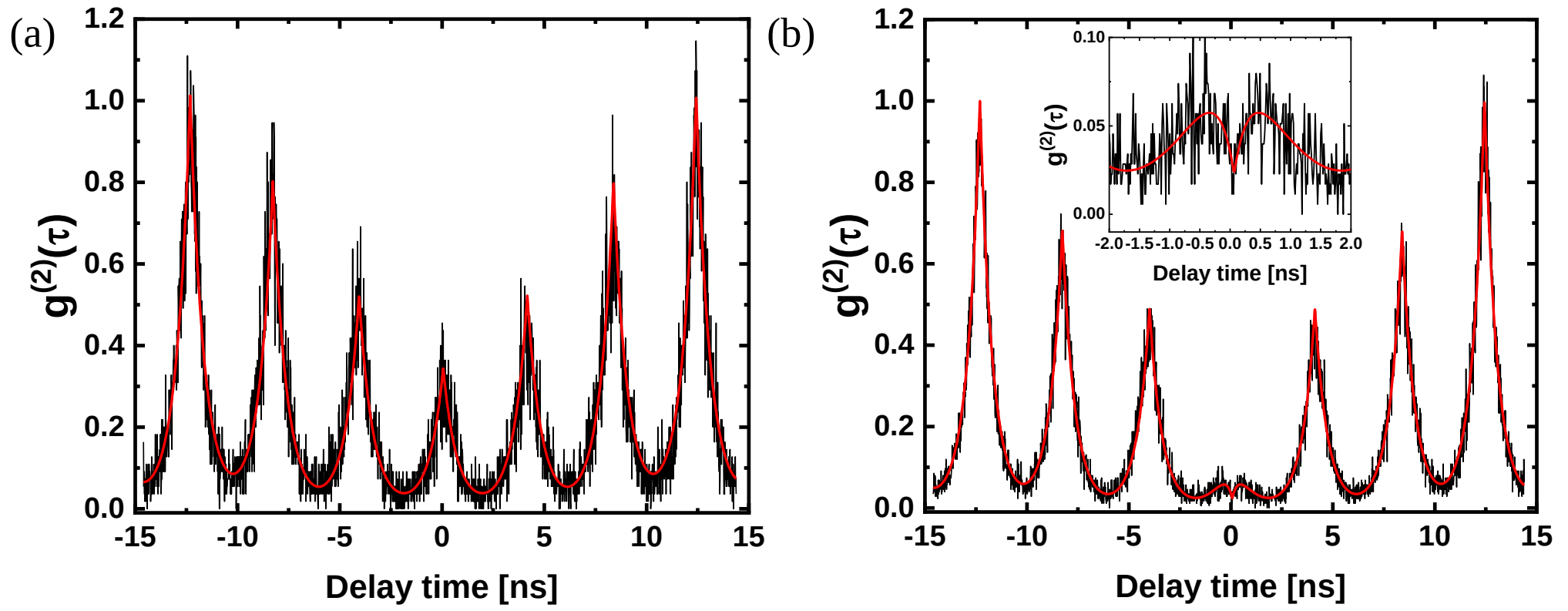

Fig. 4. The HOM interference between photons delayed by 4 ns from a single InGaAs QD reveals indistinguishability of (65.6 ± 19.1)%, with the post-selected value being (93.1 ± 0.4)%. This determination is based on a comparison of cross-polarized **(a)** and co-polarized **(b)** measurements. The data is acquired under non-resonant excitation at 930 nm, at a temperature of 4.2 K, and with a pump power of 80 µW. The solid red curve represents the fit of Eq. 2.

## 4. Conclusion

In summary, we have presented the direct growth of high-quality, InGaAs/GaAs QDs on a Si substrate emitting single photons with excellent optical and quantum optical properties. Our results were accomplished through the utilization of a GaP buffer as an intermediary layer and meticulous mitigation of defects at the III/V-Si interface. We achieved a high-quality single-photon source, characterized by a typical lifetime of $\tau = (1.54 \pm 0.01)$ ns and a high multi-photon suppression, $g^{(2)}(0) = (3.7 \pm 0.2) \cdot 10^{-2}$ (post-selected: $g^{(2)}_{Post} = (3.5 \pm 0.2) \cdot 10^{-3}$). Moreover, our investigations have yielded visibility of approximately $(65.6 \pm 19.1)\%$ (post-selected: $V^{HOM}_{Post} = (93.1 \pm 0.4)\%$ under non-resonant wetting layer excitation at 930 nm. This achievement elevates the InGaAs/GaAs QDs on Si to a comparable stature with state-of-the-art emitters grown on a GaAs substrate. It is worth noting that these QDs not only offer excellent quantum optical properties but also exhibit a high extraction efficiency of up to $(18 \pm 1)$ %. Remarkably, this performance is undeterred by the defects arising from the growth of III/V materials on the Si platform. This significant breakthrough not only highlights the excellent quantum optical properties of QDs on Si but also emphasizes the supplementary benefits of CMOS compatibility and notably reduced manufacturing expenses. The combination of remarkable quantum optical properties with the potential of Si-based electronics constitutes a noteworthy advancement in the pursuit of advanced and economically viable quantum photonics.

## Acknowledgments

The authors further acknowledge Martin Von Helversen, Aris Koulas-Simos, Lucas Rickert, and Daniel Vajner for their invaluable technical support and engaging scientific discussions, which greatly contributed to the development and success of this research.

## Disclosures

The authors declare no conflict of interest.

## Data Availability Statement

The data that support the findings of this study are available from the corresponding author upon reasonable request.

## Reference

1. B. Shi and J. Klamkin, "Defect engineering for high quality InP epitaxially grown on on-axis (001) Si," J. Appl. Phys. **127**, (2020).

2. W.-C. Kuo, H.-C. Hsieh, W. Chih-Hung, H. Wen-Hsiang, C.-C. Lee, and J.-Y. Chang, "High Quality GaAs Epilayers Grown on Si Substrate Using 100 nm Ge Buffer Layer," Int. J. Photoenergy **2016**, 1–5 (2016).

3. Z. Wang, B. Tian, M. Pantouvaki, W. Guo, P. Absil, J. Van Campenhout, C. Merckling, and D. Van Thourhout, "Room-temperature InP distributed feedback laser array directly grown on silicon," Nat. Photonics **9**, 837–842 (2015).

4. C. Shang, K. Feng, E. T. Hughes, A. Clark, M. Debnath, R. Koscica, G. Leake, J. Herman, D. Harame, P. Ludewig, Y. Wan, and J. E. Bowers, "Electrically pumped quantum-dot lasers grown on 300 mm patterned Si photonic wafers," Light Sci. Appl. **11**, 299 (2022).

5. W. L. Ng, M. A. Lourenço, R. M. Gwilliam, S. Ledain, G. Shao, and K. P. Homewood, "addendum: An efficient room-temperature silicon-based light-emitting diode," Nature **414**, 470–470 (2001).

6. R. Claps, D. Dimitropoulos, V. Raghunathan, Y. Han, and B. Jalali, "Observation of stimulated Raman amplification in silicon waveguides," Opt. Express **11**, 1731 (2003).

7. D. A. B. Miller, "Rationale and challenges for optical interconnects to electronic chips," Proc. IEEE **88**, 728–749 (2000).

8. Y. Shi, Y. Zhang, Y. Wan, Y. Yu, Y. Zhang, X. Hu, X. Xiao, H. Xu, L. Zhang, and B. Pan, "Silicon photonics for high-capacity data communications," Photonics Res. **10**, A106 (2022).

9. A. V. Krishnamoorthy, Ron Ho, Xuezhe Zheng, H. Schwetman, Jon Lexau, P. Koka, GuoLiang Li, I. Shubin, and J. E. Cunningham, "Computer Systems Based on Silicon Photonic Interconnects," Proc. IEEE **97**, 1337–1361 (2009).

10. M. Benyoucef and J. P. Reithmaier, "Direct growth of III–V quantum dots on silicon substrates: structural and optical properties," Semicond. Sci. Technol. **28**, 094004 (2013).

11. I. J. Luxmoore, R. Toro, O. Del Pozo-Zamudio, N. A. Wasley, E. A. Chekhovich, A. M. Sanchez, R. Beanland, A. M. Fox, M. S. Skolnick, H. Y. Liu, and A. I. Tartakovskii, "III–V quantum light source and cavity-QED on Silicon," Sci. Rep. **3**, 1239 (2013).

12. R. Katsumi, Y. Ota, A. Osada, T. Yamaguchi, T. Tajiri, M. Kakuda, S. Iwamoto, H. Akiyama, and Y. Arakawa, "Quantum-dot single-photon source on a CMOS silicon photonic chip integrated using transfer printing," APL Photonics **4**, 036105 (2019).

13. A. Musiał, P. Holewa, P. Wyborski, M. Syperek, A. Kors, J. P. Reithmaier, G. Sęk, and M. Benyoucef, "High-Purity Triggered Single-Photon Emission from Symmetric Single InAs/InP Quantum Dots around the Telecom C-Band Window," Adv. Quantum. Technol. **3**, 1900082 (2020).

14. S. L. Portalupi, M. Jetter, and P. Michler, "InAs quantum dots grown on metamorphic buffers as non-classical light sources at telecom C-band: a review," Semicond. Sci. Technol. **34**, 053001 (2019).

15. A. Thoma, P. Schnauber, M. Gschrey, M. Seifried, J. Wolters, J.-H. Schulze, A. Strittmatter, S. Rodt, A. Carmele, A. Knorr, T. Heindel, and S. Reitzenstein, "Exploring Dephasing of a Solid-State Quantum Emitter via Time- and Temperature-Dependent Hong-Ou-Mandel Experiments," Phys. Rev. Lett. **116**, 033601 (2016).

16. P. Senellart, G. Solomon, and A. White, "High-performance semiconductor quantum-dot single-photon sources," Nat. Nanotechnol. **12**, 1026–1039 (2017).

17. T. Heindel, J.-H. Kim, N. Gregersen, A. Rastelli, and S. Reitzenstein, "Quantum dots for photonic quantum information technology," Adv. Opt. Photonics **15**, 613 (2023).

18. A. Blais, S. M. Girvin, and W. D. Oliver, "Quantum information processing and quantum optics with circuit quantum electrodynamics," Nat. Phys. **16**, 247–256 (2020).

19. D. A. Vajner, L. Rickert, T. Gao, K. Kaymazlar, and T. Heindel, "Quantum Communication Using Semiconductor Quantum Dots," Adv. Quantum. Technol. **5**, 2100116 (2022).

20. H. Wang, J. Qin, X. Ding, M.-C. Chen, S. Chen, X. You, Y.-M. He, X. Jiang, L. You, Z. Wang, C. Schneider, J. J. Renema, S. Höfling, C.-Y. Lu, and J.-W. Pan, "Boson Sampling with 20 Input Photons and a 60-Mode Interferometer at $10^{14}$-Dimensional Hilbert Space," Phys. Rev. Lett. **123**, 250503 (2019).

21. A. Strittmatter, A. Holzbecher, A. Schliwa, J.-H. Schulze, D. Quandt, T. D. Germann, A. Dreismann, O. Hitzemann, E. Stock, I. A. Ostapenko, S. Rodt, W. Unrau, U. W. Pohl, A. Hoffmann, D. Bimberg, and V. Haisler, "Site-controlled quantum dot growth on buried oxide stressor layers," Phys. Status Solidi A **209**, 2411–2420 (2012).

22. C. Schneider, A. Huggenberger, M. Gschrey, P. Gold, S. Rodt, A. Forchel, S. Reitzenstein, S. Höfling, and M. Kamp, "In(Ga)As/GaAs site-controlled quantum dots with tailored morphology and high optical quality," Phys. Status Solidi A **209**, 2379–2386 (2012).

23. S. Rodt and S. Reitzenstein, "Integrated nanophotonics for the development of fully functional quantum circuits based on on-demand single-photon emitters," APL Photonics **6**, (2021).

24. A. Kaganskiy, M. Gschrey, A. Schlehahn, R. Schmidt, J.-H. Schulze, T. Heindel, A. Strittmatter, S. Rodt, and S. Reitzenstein, "Advanced *in-situ* electron-beam lithography for deterministic nanophotonic device processing," Rev. Sci. Instrum. **86**, (2015).

25. K. Volz, A. Beyer, W. Witte, J. Ohlmann, I. Németh, B. Kunert, and W. Stolz, "GaP-nucleation on exact Si (001) substrates for III/V device integration," J. Cryst. Growth **315**, 37–47 (2011).

26. A. Beyer, B. Haas, K. I. Gries, K. Werner, M. Luysberg, W. Stolz, and K. Volz, "Atomic structure of (110) anti-phase boundaries in GaP on Si(001)," Appl. Phys. Lett. **103**, (2013).

27. T. Miyazawa, K. Takemoto, Y. Nambu, S. Miki, T. Yamashita, H. Terai, M. Fujiwara, M. Sasaki, Y. Sakuma, M. Takatsu, T. Yamamoto, and Y. Arakawa, "Single-photon emission at 1.5 μm from an InAs/InP quantum dot with highly suppressed multi-photon emission probabilities," Appl. Phys. Lett. **109**, (2016).

28. P. A. Dalgarno, J. McFarlane, D. Brunner, R. W. Lambert, B. D. Gerardot, R. J. Warburton, K. Karrai, A. Badolato, and P. M. Petroff, "Hole recapture limited single photon generation from a single n-type charge-tunable quantum dot," Appl. Phys. Lett. **92**, (2008).

29. S. Fischbach, A. Schlehahn, A. Thoma, N. Srocka, T. Gissibl, S. Ristok, S. Thiele, A. Kaganskiy, A. Strittmatter, T. Heindel, S. Rodt, A. Herkommer, H. Giessen, and S. Reitzenstein, "Single Quantum Dot with Microlens and 3D-Printed Micro-objective as Integrated Bright Single-Photon Source," ACS Photonics **4**, 1327–1332 (2017).

30. L. Zhai, M. C. Löbl, G. N. Nguyen, J. Ritzmann, A. Javadi, C. Spinnler, A. D. Wieck, A. Ludwig, and R. J. Warburton, "Low-noise GaAs quantum dots for quantum photonics," Nat. Commun. **11**, 4745 (2020).

# High-quality single InGaAs/GaAs quantum dot growth on a CMOS-compatible silicon substrate for quantum photonic applications: Supplemental document

**Imad Limame,[1] Peter Ludewig,[2] Ching-Wen Shih,[1] Marcel Hohn,[1] Chirag C. Palekar,[1] Wolfgang Stolz,[2] AND Stephan Reitzenstein[1,*]**

[1]*Institute for Solid State Physics, Technical University of Berlin, Hardenbergstraße 36, D-10623 Berlin, Germany*
[2]*Structure and Technology Research Laboratory, Philipps University of Marburg, Hans-Meerwein-Straße 6, D-35032 Marburg, Germany*
[*]*stephan.reitzenstein@physik.tu-berlin.de*

This supplementary materials document provides additional information on the surface analyses of the quantum dot (QD) samples, on the experimental setup and on the spectroscopic study a of single QD.

## 1. Morphological study

Fig. S1. displays the surface morphology of the studied sample, which was grown on a silicon substrate. The surface was characterized using an atomic force microscope (AFM) in (a) and a optical microscope in (b). In both cases, we clearly observe surface roughness that is not typically expected for growth on a GaAs substrate. The coalescence of GaP leads to the formation of stacking faults, as illustrated in the 10 µm x 10 µm AFM image, Fig. S1(a). Furthermore, we have observed the existence of misfit dislocations, which appear as lateral lines seen in Fig. S1. (b). These dislocations form at the Si/GaP interface to compensate for the lattice mismatch.

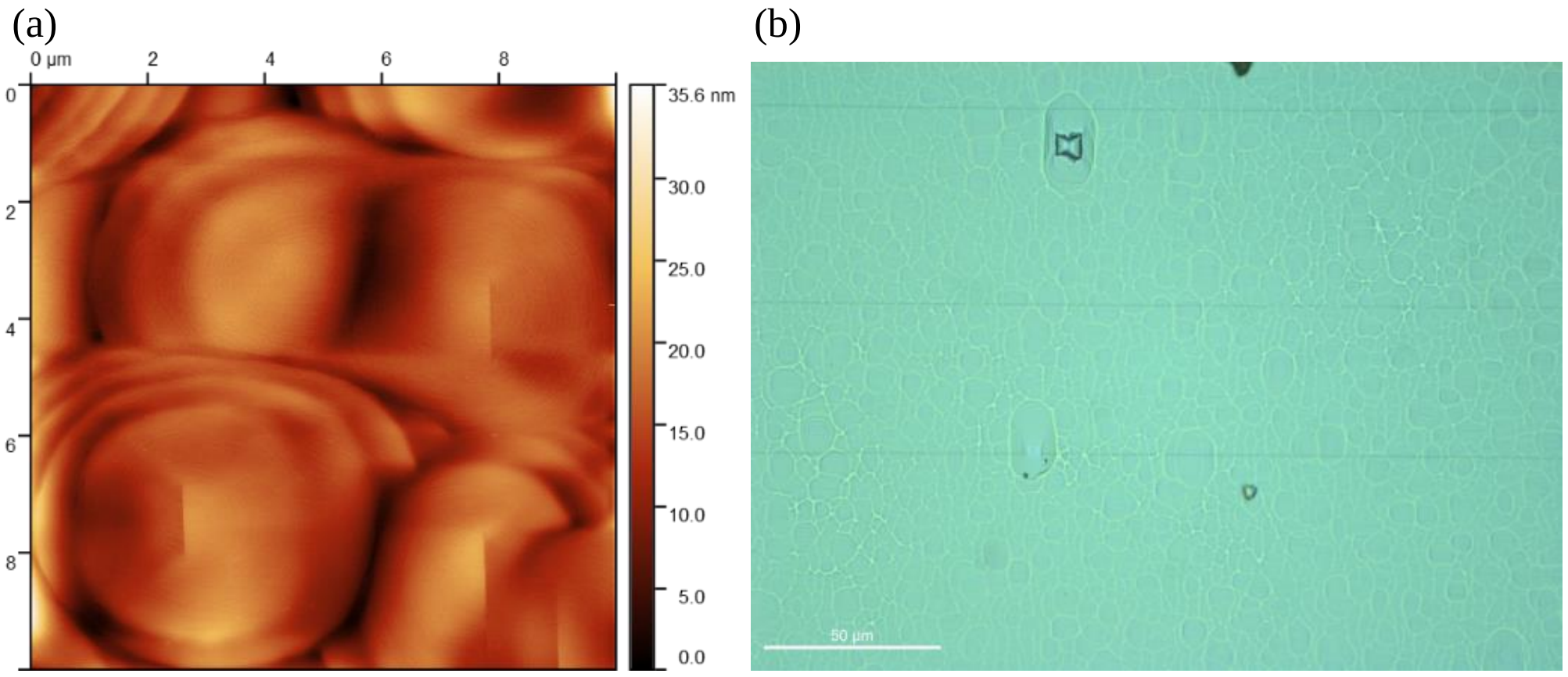

Fig. S1. (a) 10 µm x 10 µm AFM image of the morphology of the final structure revealing a rough surface, comprising grain-shaped islands between 6 and 12 µm in lateral size and up to 40 nm in height emerging from the GaP nucleation on the Si substrate. (b) Optical image of misfit dislocations lines throughout the sample surface.

## 2. Optical setup

A standard micro-photolumenescence (µPL) setup was used to conduct the optical characterization of the grown sample. Fig. S2 shows a shematic illustration of the µPL setup comprising a close-cycle cryostat with a three-axis piezo stage, the pulsed laser, the monochromator with a 1200 lines/mm grating, and a Si CCD. The Hanbury Brown and Twiss (HBT) setup is based on a single-mode fiber beamsplitter and uses two channels of superconducting nanowire single-photon detectors (SNSPDs) to detect the emitted photons and a time-correlated single-photon counter (TCSPC) to measure the auto-correlation function ($g^{(2)}(\tau)$). The Hong-Ou-Mandel (HOM) interference effect is measured with a 4 ns time delay. The emitter light from the InGaAs/GaAs QDs on Si is equally split using a combination of a λ/2 plate and a polarization beamsplitter (PBS), afterward two λ/2 and λ/4 plates are used to precisely control the polarization of each path.

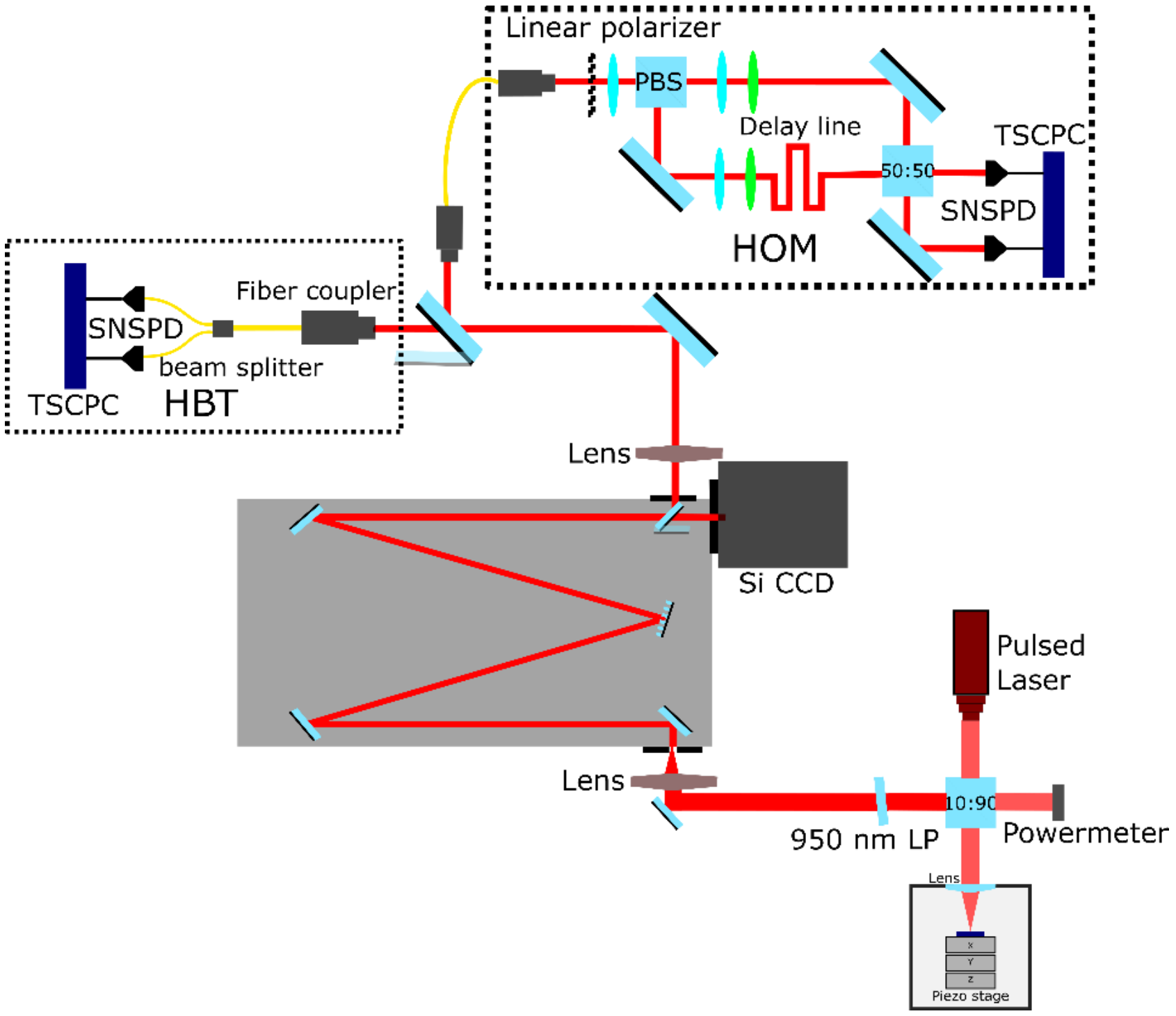

Fig. S2. A schematic illustration of a standard µPL setup used for the characterization of the optical and quantum optical properties of the InGaAs QDs grown on a silicon wafer.

## 3. Optical characterization and FEM simulation

In Fig. S3 (a) the integrated µPL intensity as a function of the excitation power is ploted in a double logarithmic representation. The data points represent the measured intensity, while the solid line depicts the power-law dependence of I on P for the exciton, biexciton, and charged biexciton in black with a slope of $1.15 \pm 0.11$, red with a quadratic dependency and blue with $2.48 \pm 0.12$, respectively. In (b) Polarization-resolved measurement of the three QD transitions showing a clear exciton/biexciton behavior and a fine structure splitting of $\Delta_{FSS} = 30$ µeV, while the charged state shows no reaction to the polarization angle.

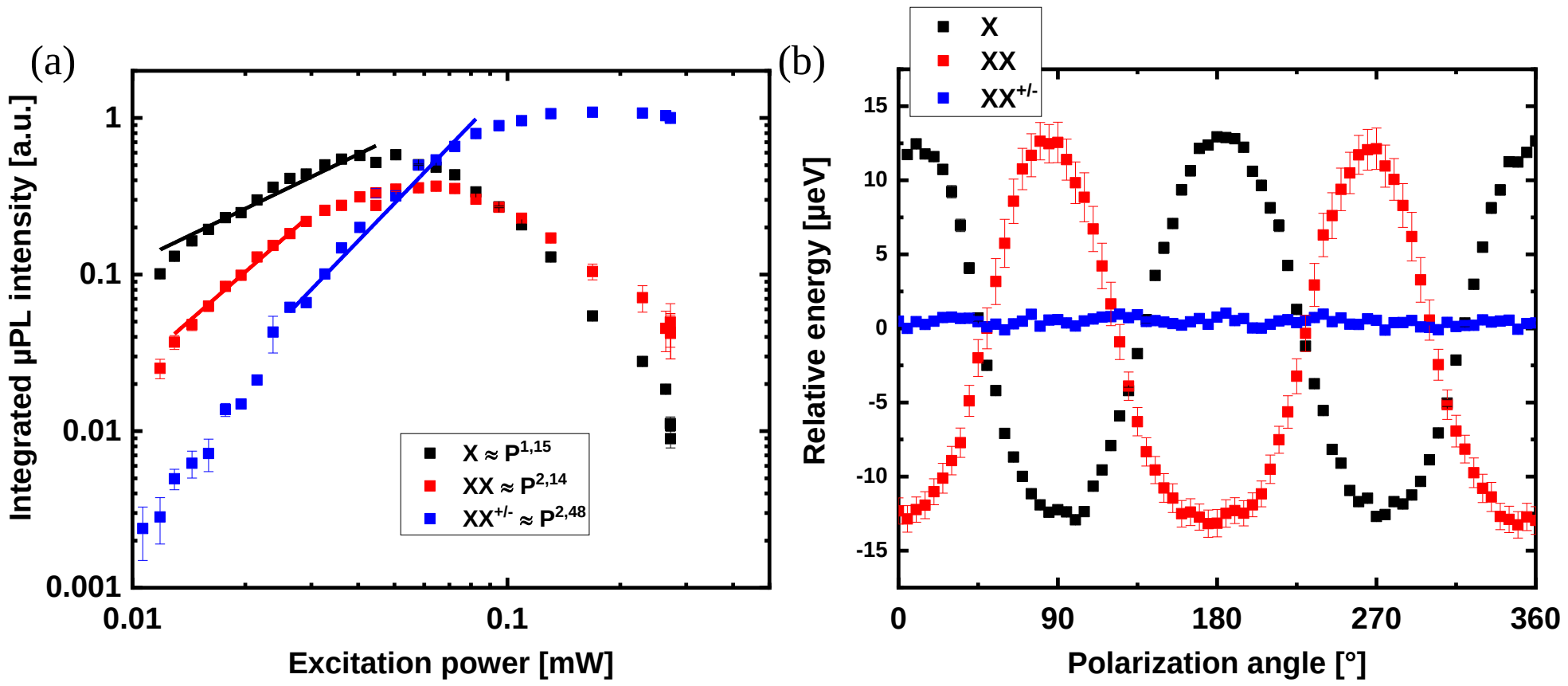

Fig. S3. (a) Double logarithmic plot of the integrated µPL intensity as a function of the excitation power. (b) Polarization-resolved measurement reveal a fine structure splitting of $\Delta_{FSS}$= 30 µeV.

Finite element method (FEM) simulations of the studied structure were conducted to estimate the upper limit of the extraction effiency as a function of the emission wavelength, resulting in the curves in Fig. S4. (a). (b) showcases the measured µPL intensity of a bright emitter as a function of the excitation power, which was evoluted for the extraction effiency of $(18 \pm 1)$ %. In the inset, the spectrum of the investigated QD at saturation under pulsed excitation with a laser repetition frequency of 80 MHz.

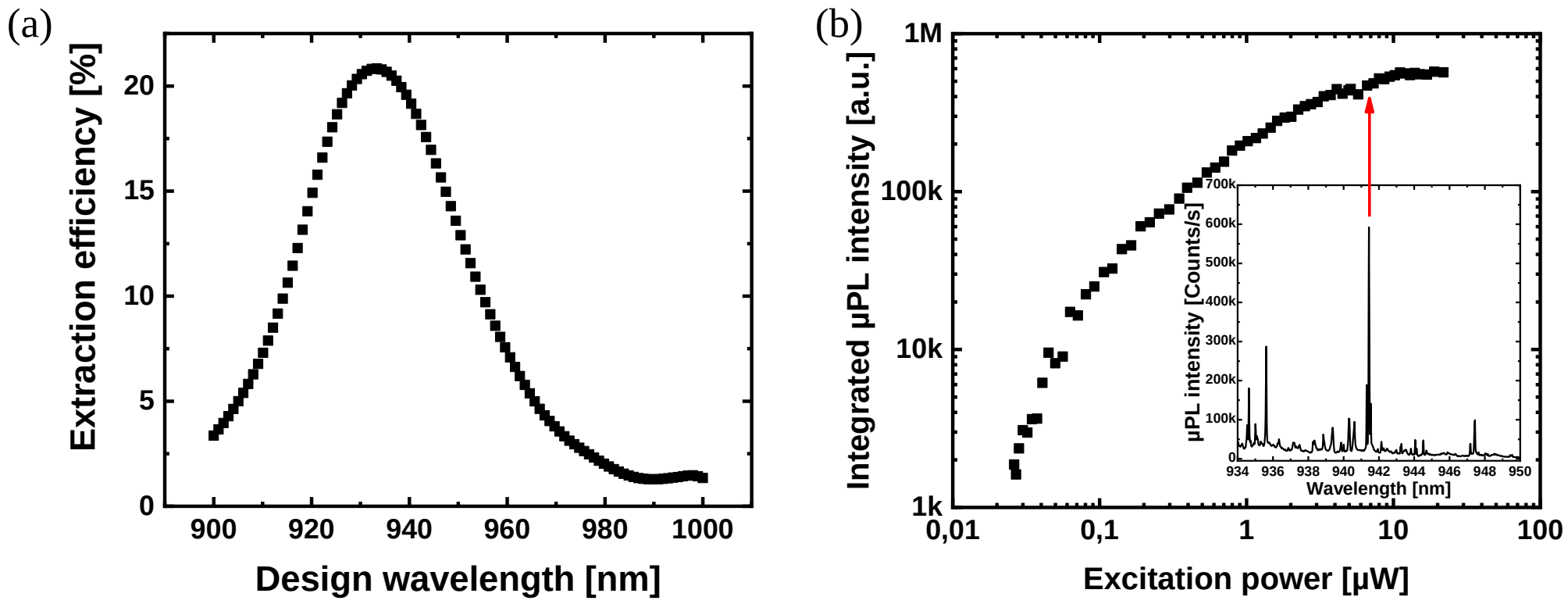

Fig. S4. (a) Simulated extraction efficiency are illustrated as a function of the wavelength in black and red, respectively. **(b)** The measured µPL intensity as a function of the excitation power. In the inset, the spectrum of the investigated QD at saturation.